# Formation of single-phase disordered Cs$_x$Fe$_{2-y}$Se$_2$ at high pressure

V. Svitlyk[1*], E. Pomjakushina[2], A. Krzton-Maziopa[3], K. Conder[2] and M. Mezouar[1]

[1]ID27 High Pressure Beamline, European Synchrotron Radiation Facility, 38000 Grenoble, France
[2] Laboratory for Multiscale Materials Experiments, Paul Scherrer Institute, 5232 Villigen, Switzerland
[3]Warsaw University of Technology, Faculty of Chemistry, 00-664 Warsaw, Poland

*e-mail: svitlyk@esrf.fr

**Abstract**
A single-phase high pressure (HP) modification of Cs$_x$Fe$_{2-y}$Se$_2$ was synthesized at 11.8 GPa at ambient temperature. Structurally this polymorph is similar to the minor low pressure (LP) superconducting phase, namely they both crystallize in a ThCr$_2$Si$_2$-type structure without ordering of the Fe vacancies within the Fe-deficient FeSe$_4$ layers. The HP Cs$_x$Fe$_{2-y}$Se$_2$ polymorph is found to be less crystalline and nearly twice as soft compared to the parent major and minor phases of Cs$_x$Fe$_{2-y}$Se$_2$. It can be quenched to low pressures and is stable at least on the scale of weeks. At ambient pressure the HP polymorph of Cs$_x$Fe$_{2-y}$Se$_2$ is expected to exhibit different superconducting properties compared to its LP minor phase ($T_c$ = 27 K).

**Introduction**
The phenomena of superconductivity was discovered more than a century ago (Hg with a $T_c$ = 4 K) [1], but to date no material exhibiting a superconducting response at ambient temperature has been reported. While the recently discovered FeSe phase also possesses a low critical temperature ($T_c$ = 8 K [2]), there exist a number of ways to enhance its superconducting performance. For instance, the transition temperature drastically increases for FeSe in a monolayer form ($T_c$ > 100 K [3]). In addition, its $T_c$ can be enhanced by intercalation: for the Fe-deficient FeSe systems intercalated by alkali metals (A), the $T_c$ was reported to reach 30 K [4-6]. Further increase in the $T_c$ of $A_x$Fe$_{2-y}$Se$_2$ can be achieved through a control of Fe occupancies. Phases with $T_c$ exceeding 40 K can be achieved through specific synthetic procedures, including precise control of stoichiometry and annealing conditions, and are characterized by fully, or close to fully, occupied Fe sites [7-9]. In addition, their formation can be mediated by NH$_3$ molecules [10-13].

Application of external pressure is yet another tool that allows control of physical properties, including FeSe-based superconducting materials. For the FeSe phase itself, high-pressure (HP) induces more than a four-fold increase in its critical temperature ($T_c$ = 36 K around 7 GPa [14]). Above 7 GPa, structural transformation of FeSe into a topologically-different polymorph suppresses superconductivity and results in the a formation of the famous superconducting dome, with a complex phase composition [15]. In contrast, the $T_c$ of the intercalated $A_x$Fe$_{2-y}$Se$_2$ (A – alkali metals) decreases upon application of external pressure, and eventually superconductivity vanishes [16,17]. Upon further pressure increase (P > 11.5 GPa) a new superconducting phase (denoted as SCII) with a $T_c$ reaching 48 K was reported [18]. This result, however, has not yet been experimentally reproduced with independent studies reporting much lower critical temperatures (~5 or ~20 K depending on sample preparation procedures [19,20]).

The structural properties of alkali-intercalated FeSe phases are complex, as are the ones of the parent FeSe phase [15]. Firstly, the $A_x$Fe$_{2-y}$Se$_2$ family features intrinsic phase separation and the second minor phase is responsible for the observed superconductivity [21]. This phase has a nominal composition of $A_{0.5}$Fe$_{2.0}$Se$_2$ [22-24] and features a symmetry not higher than monoclinic [25]. The phase separation is suppressed with temperature [22] but this process is kinetically inhibited with pressure [26]. The main phase of $A_x$Fe$_{2-y}$Se$_2$ is deficient both on A and Fe sites with a typical composition of $A_{0.8}$Fe$_{1.6}$Se$_2$ [21]. At ambient conditions the Fe-vacancies of the main phase are ordered resulting in a formation of a $\sqrt{5}\times\sqrt{5}$ superstructure [27-31] ($I4/m$ symmetry). This ordering can be suppressed with temperature and pressure with a resulting symmetry of $I4/mmm$ [26,27,29].

Similar to the behavior of phase separation, the pressure-dependent order-disorder transition is also kinetically inhibited.

Available structural data on the SCII phases of $A_x$Fe$_{2-y}$Se$_2$ are not detailed, and sometimes even contradicting. These phases were reported to preserve tetragonal symmetry [18,19] following a collapse in the crystallographic *c* direction of the parent phases (tetragonal to collapsed tetragonal phase transition [19]). In addition, a sudden decrease in the Fe–Fe distances was observed during this transformation [32]. Independent studies point to either phase-separated or phase-merged composition of the SCII region [19,32]. In this paper we report the first diffraction studies on monocrystalline superconducting Cs$_x$Fe$_{2-y}$Se$_2$ as a function of pressure and temperature (up to 20 GPa and down to 20 K). The use of single crystal diffraction using synchrotron radiation allowed us to track in detail the mechanism of formation of the SCII phase in the $A_x$Fe$_{2-y}$Se$_2$ family. We show that the SCII phase is formed directly from the minor superconducting phase of the parent $A_x$Fe$_{2-y}$Se$_2$ sample at the expense of the main non-superconducting phase. The SCII region is, therefore, composed of one single phase and, at least for the case of the Cs$_x$Fe$_{2-y}$Se$_2$ system, this HP state can be quenched to low pressures.

**Experiment**

Single crystals of Cs$_x$Fe$_{2-y}$Se$_2$ studied in this work are the same as in our previous publications [27,29]. They were grown by the Bridgman method and are superconducting at $T_c$ = 27 K [5]. Their composition, as established from single crystal synchrotron radiation diffraction, is Cs$_{0.83(1)}$Fe$_{1.71(1)}$Se$_2$ [27] (corresponds to a composition of the main phase). Elemental composition of the cleaved crystals obtained from X-ray fluorescence spectroscopy is equal to Cs$_{0.74}$Fe$_{1.54}$Se$_2$ (2% accuracy) [29]. This formula corresponds to the average composition of the crystals, e.g. main and minor phases were measured simultaneously.

Single crystal diffraction data as a function of pressure were collected at the ID27 High Pressure Beamline at the European Synchrotron Radiation Facility (Grenoble, France) at room temperature (RT) and 20 K. For each run pressures up to 20 GPa were generated by diamond anvil cells (DAC) with 600 µm diamond culets. Samples in a shape of plates with typical dimensions of 20x20x10 µm were contained in stainless steel gaskets with holes of 300 and thickness of 90 µm. He gas was used as a pressure transmitting medium in order to ensure highly hydrostatic conditions [33] and the pressure was measured using a ruby fluorescence technique [34]. Low temperature was achieved using the custom in-house He flow cryostat. Synchrotron radiation was ($\lambda$ = 0.3738 Å) focused to a spot size of 3x3 µm (FWHM) at the sample. During data collection samples were rotated by 60 degrees in continuous (panoramic images) or 1 degree slicing (3D reconstruction of reciprocal spaces) modes. These data were recorded on a flat panel PerkinElmer detector and the experimental slices of reciprocal space were generated using the CrysAlisPro package [35]. All experimentally available reciprocal space was examined but only the most informative parts have been included into the main part of the manuscript. Additional slices are available in the Supplementary Material [36].

**Results and discussion**

Fine structural features of the studied Cs$_x$Fe$_{2-y}$Se$_2$ sample - namely $\sqrt{5}$x$\sqrt{5}$ superstructure reflections indicative of the Fe-vacancies ordering of the main phase (Fig. 1, left, *hk*0 reconstruction of the reciprocal layer at 0.2 GPa), and the phase separation (Fig. 2, left, enlargement of a region containing 022 reflections) - can be clearly observed and traced as a function of pressure within our experimental single crystal diffraction data. Firstly, upon application of HP peaks of the main and secondary phases approach and eventually merge into one phase at 11.8 GPa (Fig. 2, top and middle, Fig. 3, top, left). The $\sqrt{5}$x$\sqrt{5}$ superstructure reflections of the main phase and diffuse rods of the secondary phase [25] disappear at the same pressure of 11.8 GPa (within a step resolution of 0.3 GPa, Fig. 2, bottom, Fig. 3, top, right). Therefore, 11.8 GPa corresponds to a formation of a new phase with disorder within the Fe-sublattice, *i.e.* of *I*4/*mmm* symmetry, and of a single-phase nature (Fig. 1, right, *hk*0 reconstruction of the 13.2 GPa data is shown). In addition, the crystallinity of the sample significantly reduced after the transition (Fig. 1). We note, however, that the absence of the $\sqrt{5} \times \sqrt{5}$ superstructure reflections in HP $Cs_xFe_{2-y}Se_2$ is not caused by the HP amorphization but is an intrinsic structural property of this phase. Indeed, the superstructure reflections lose already 97% of their intensity before the transition at 11.8 GPa (Fig. 3, top, right), *i.e.* before the sudden decrease in crystallinity of $Cs_xFe_{2-y}Se_2$.

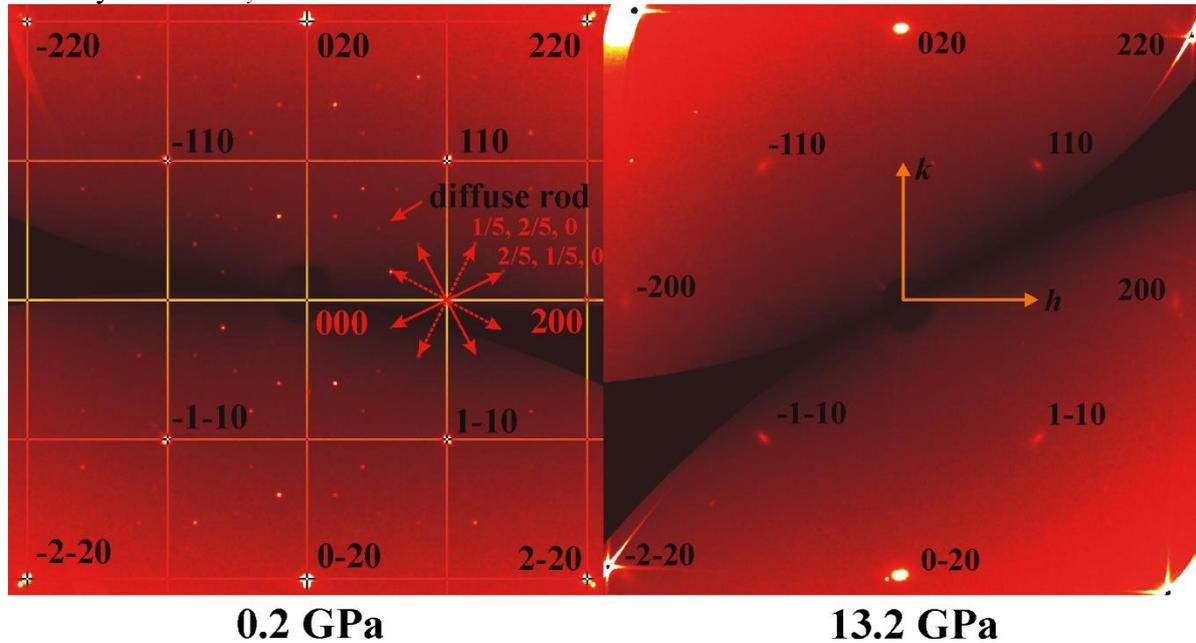

Figure 1. Reconstruction of the *hk*0 reciprocal layers of $Cs_xFe_{2-y}Se_2$ at 0.2 (left) and 13.2 GPa (right). Left: yellow grid corresponds to a reciprocal lattice of the average *I*4/*mmm* structure of $Cs_xFe_{2-y}Se_2$; a star indicates a group of $\sqrt{5} \times \sqrt{5}$ superstructure reflections (solid and dash lines correspond to two different twin domains); a separate arrow indicates a slice through a diffuse rod of the minor phase [25]. Right: yellow arrows mark a new lattice of the HP $Cs_xFe_{2-y}Se_2$ structure (*I*4/*mmm* symmetry). Additional reflections visible at 13.2 GPa (right) are not commensurate neither with the main nor the minor phases of $Cs_xFe_{2-y}Se_2$, and originate from the sample environment (diamonds, solid He/ruby crystals).

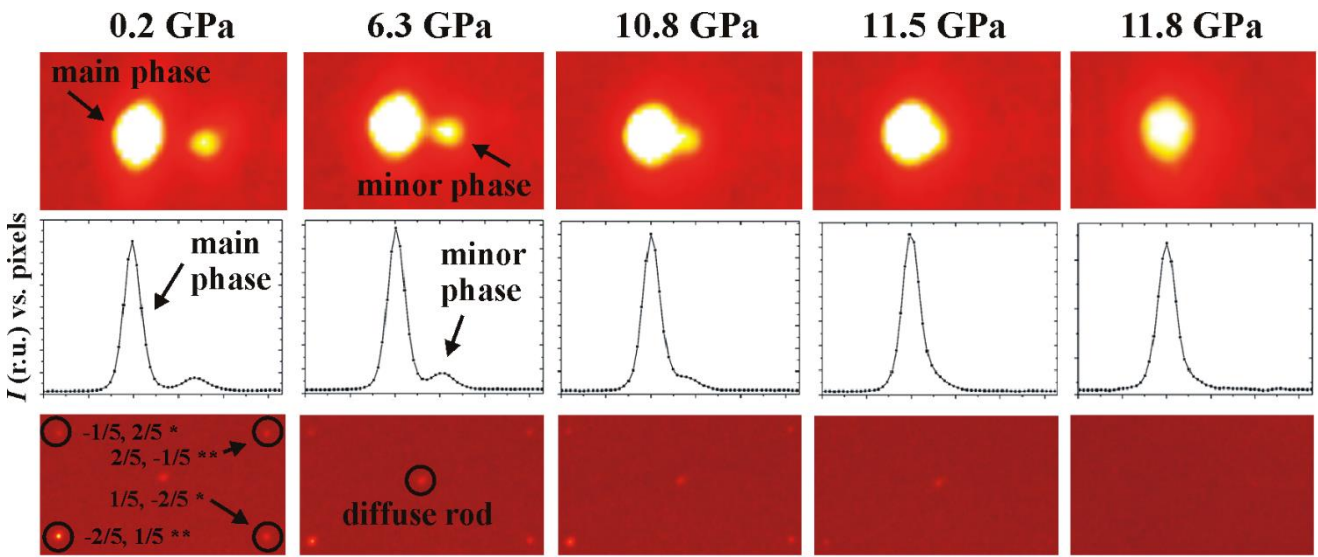

Figure 2. Evolution of fine structural features of $Cs_xFe_{2-y}Se_2$ as a function of pressure. Top: high resolution zoom on a region of the reciprocal space containing 020 reflections of the main and minor phases; positions of the reflections of the main phase were kept fixed as a reference. Middle:

corresponding profiles of the 020 reflections of the main and minor phases. Bottom: $\sqrt{5}\text{x}\sqrt{5}$ superstructure reflections of the main phase and a slice through the diffuse rod of the minor phase.

Quantitative information on the behavior of unit cell parameters of the major and secondary minor phases is shown in Fig. 3 (bottom). The $a$ parameter of the main $I4/m$ phase (black curve) follows a uniform compression until 11 GPa, then suddenly collapses and merges with the corresponding parameter of the minor phase (red line). Therefore at this pressure the initially minor superconducting phase becomes the major single phase, and continues to follow its initial low pressure (LP) compression curve (red curve which transforms into black one at 11.8 GPa). During the experiment orientation of the studied single crystal plate in the DAC was not favorable to reliably extract $d$-spacing along the $c$ axis of the minor phase. So only the behavior of the $c$ parameter of the main phase is shown in Fig. 3 (bottom, right, blue curve). Similarly to the $a$ parameter (Fig. 3, bottom, left), it indicates a collapse into a monophasic state at 11.8 GPa and the corresponding unit cell volume exhibits analogous behavior (Fig. 3, bottom, right, red curve). Reflections containing contributions from the $c$ direction (Fig. 3, bottom, right, 03-3 reflection is shown) however feature a rapid amorphization after the transition at 11.8 GPa and, as a result, a corresponding quantitative behavior of the $c$ parameter after the transition and upon decompression could not be reliably extracted. Clearly, behavior of structural parameters around 11.8 GPa (Fig. 3) is indicative of a first-order structural transformation at this pressure.

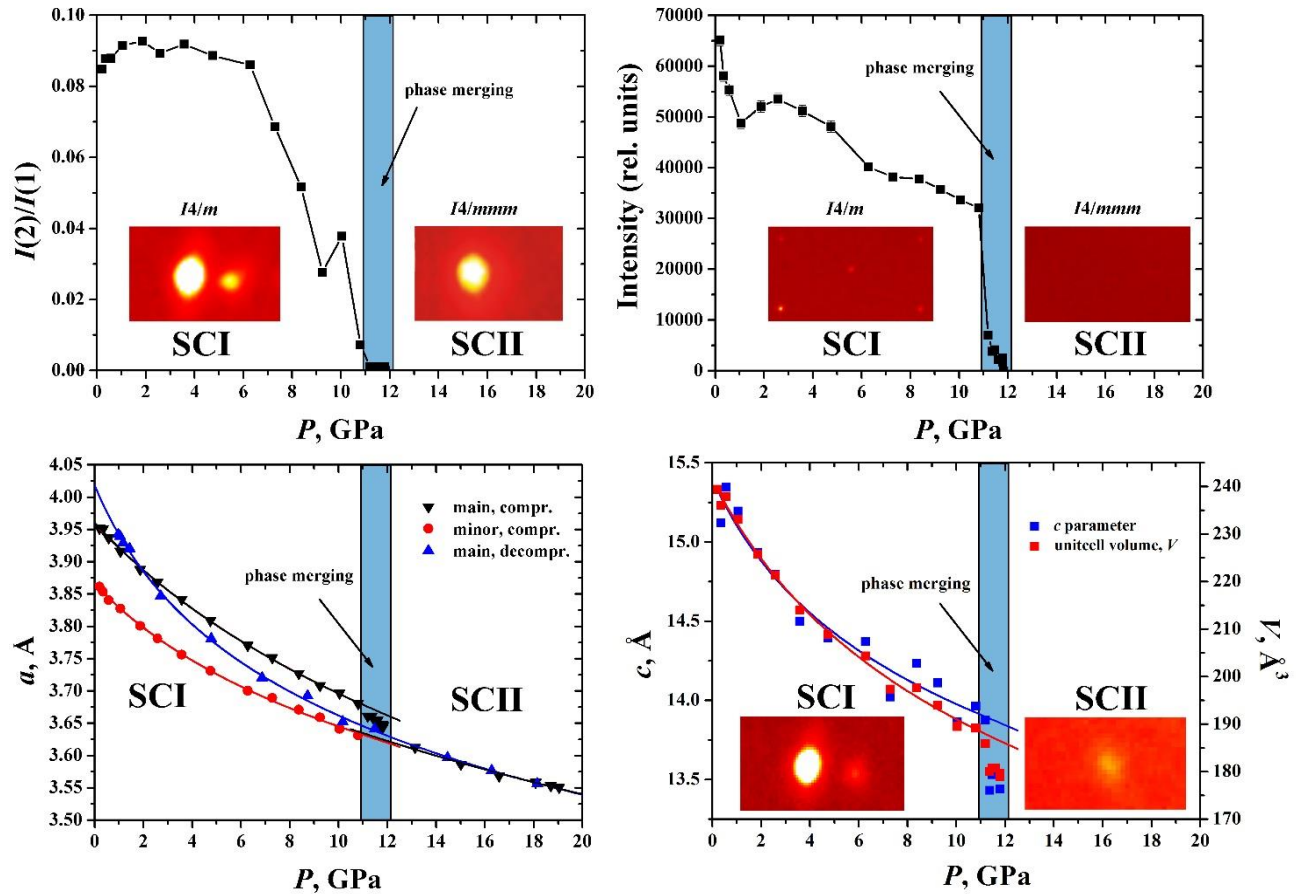

Figure 3. Evolution of fine structural parameters of $Cs_xFe_{2-y}Se_2$ as a function of pressure. Top left: intensity ratio of the minor and main phases; insets show the 020 reflections. Top right: intensity of $\sqrt{5}\text{x}\sqrt{5}$ superstructure reflections. Bottom left: experimental $a$ parameters as directly refined from the behavior of the 020 reflection of the main and secondary phases (Fig. 2). Bottom right: $c$ parameter (extracted from the pair of the $03\bar{3}$ and 020 reflections) and unit-cell volume of the main phase; insets show the $03\bar{3}$ reflections.

Merging of the LP minor and LP main phases, and a formation of a single phase sample at HP, implies diffusion of Cs and Fe atoms on a micro-meter length scale [37]. Limited data coverage intrinsic to HP experiments, multiplied by a pressure-induced reduction in crystallinity of $Cs_xFe_{2-y}Se_2$, precluded reliable refinement of occupancies of Cs and Fe atoms in this phase as a function of pressure. Nevertheless, composition of the main phase of the analogues crystal isolated from the same growth batch was found to be equal to $Cs_{0.83(1)}Fe_{1.71(1)}Se_2$ from our previous single crystal synchrotron diffraction experiment at ambient conditions [27] and equal to $Cs_{0.74}Fe_{1.54}Se_2$ (2% accuracy) [29] from X-ray fluorescence spectroscopy. This composition corresponds to the average composition of the crystal since two phases were measured simultaneously and, therefore, also reflects composition of the HP monophasic state. An evolution in composition as a function of temperature was also observed by us for a related $Rb_xFe_{2-y}Se_2$ phase where analogous phase separation is also suppressed as a function of temperature [22].

Structurally the HP modification of $Cs_xFe_{2-y}Se_2$ is very similar to the minor LP superconducting one. Specifically, both do not feature order within the Fe-deficient sublattices, i.e. the $\sqrt{5}\times\sqrt{5}$ superstructure reflections are absent, and a corresponding average symmetry is $I4/mmm$. A possible symmetry lowering, reported by us for the minor phase at ambient conditions [25], was not observed in the current experiment. Vanishing of diffuse rods that are originally present in the LP minor phase, and which originate from Cs ordering within the Cs-deficient layers [25], indicates absence of the corresponding correlations in HP $Cs_xFe_{2-y}Se_2$. In addition a general decrease in crystallinity is also evident from a change in the shape of the experimental Bragg reflections, especially along the $c$ direction, where they become more diffuse as a function of pressure (Fig. 3, bottom, right). The corresponding average model of the HP phase is represented on the Fig. 4 and corresponds to the $I4/mmm$ ThCr$_2$Si$_2$-type structure.

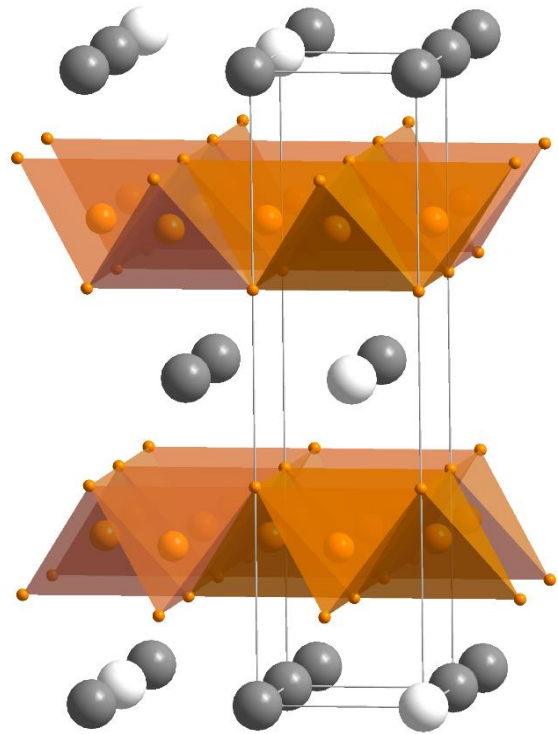

Figure 4. Average model of the HP $Cs_xFe_{2-y}Se_2$ phase corresponding to the ThCr$_2$Si$_2$-type arrangement ($I4/mmm$). Fe in the layers of the edge-shared FeSe$_4$ tetrahedra (orange) are ¾ occupied; intercalated Cs atoms (dark grey) also occupy about ¾ of available positions (Cs vacancies are highlighted).

Surprisingly, upon a decompression from a maximum achieved pressure of 19 GPa (RT) the HP modification of $Cs_xFe_{2-y}Se_2$ remained thermodynamically stable at 0.7 GPa for at least 10 days (Fig. 5). Namely, the sample remained monophasic and the $\sqrt{5}\times\sqrt{5}$ superstructure reflections indicative of the Fe-vacancies ordering did not re-appear. Interestingly, the HP modification of $Cs_xFe_{2-y}Se_2$ follows a distinct decompression path (Fig. 3, bottom, left, blue curve) as compared to the compression behavior of the parent minor and main phases (Fig. 3, bottom, left, red and black curves). Indeed, it is nearly twice as soft than the parent phases and, in addition, features larger cell parameters (at least *a* and *b*) at ambient conditions as compared to the initial ones (Table 1). The softening is likely related to a reduced crystallinity observed on diffraction data discussed above.

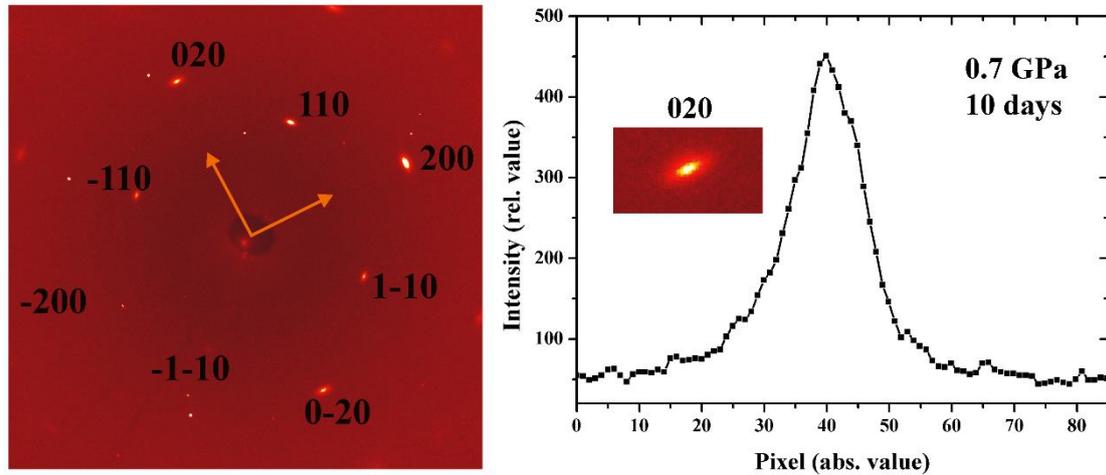

Figure 5. Left: panoramic projection of the $Cs_xFe_{2-y}Se_2$ single crystal kept at 0.7 GPa for 10 days. Right: profile of the 020 reflection highlighting the single phase nature of the sample. Similar to Figure 1, additional reflections visible in data taken at 0.7 GPa (left) are not commensurate neither with the main (including $\sqrt{5} \times \sqrt{5}$ superstructure reflections) nor the minor phases of $Cs_xFe_{2-y}Se_2$ and originate from the sample environment.

Table 1. Compressibilities of different phases of $Cs_xFe_{2-y}Se_2$ along the *a-b* unit cell directions as obtained from a Murnaghan equation of state ($V(P) = a_0(1 + B_0'\frac{P}{B_0})^{-\frac{1}{B_0'}}$, where $a_0$ is the *a* unit cell parameter at zero pressure, $B_0$ is the bulk modulus and $B_0'$ is the first pressure derivative of the bulk modulus)

| $Cs_xFe_{2-y}Se_2$ Phase | *P* range, GPa | Symmetry | $a_0$, Å | $B_0$, GPa | $B_0'$, GPa |
|---|---|---|---|---|---|
| Main LP | < 11 | *I*4/*m* | 3.960(2) | 97(3) | 11.0(8) |
| Minor LP | < 11 | *I*4/*mmm* | 3.866(2) | 95(5) | 18(1) |
| Main* HP | > 12 | *I*4/*mmm* | 3.94(5) | 60(17) | 18 (fixed) |
| Minor LP + Main* HP | 0.2 - 19 | *I*4/*mmm* | 3.866(2) | 98(3) | 17.4(6) |
| Main* decompression | 19 – 0.7 | *I*4/*mmm* | 4.019(7) | 43(4) | 17.3(8) |

*although at HP and after the decompression the sample is single-phase, for a convenience we will continue to designate the HP phase of $Cs_xFe_{2-y}Se_2$ as a main

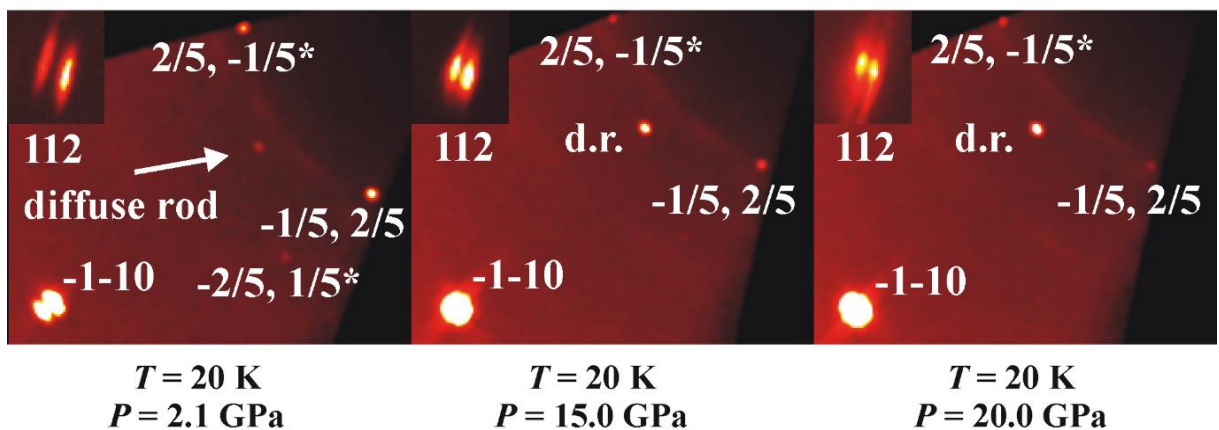

Figure 6. Evolution of $Cs_xFe_{2-y}Se_2$ with pressure at 20 K.

Finally, a $Cs_xFe_{2-y}Se_2$ crystal was compressed at 20 K (Fig. 6) in order to study temperature effects on the kinetics of the observed phenomena at room temperature, if any. Surprisingly at 20 K, where the minor phase of $Cs_xFe_{2-y}Se_2$ is superconducting both the superstructure reflections of the main phase and the two-phase state of the sample persist up to 20 GPa. However, a clear tendency towards a monophasic state analogous to the one present at room temperature is observed: (i) Bragg reflections of minor and main phases start to approach (Fig. 6, inset with a 112 peak is shown as an example) and (ii) intensities of the $\sqrt{5} \times \sqrt{5}$ superstructure reflections start to diminish. A further increase in pressure is required to complete a transition into the HP polymorph. However, it is not excluded that the corresponding transition mechanism at 20 K differs from that at room temperature. In particular, suppression of superstructure reflections and the phase merging can happen at different pressures, analogous to the temperature-dependent transition in related $Rb_xFe_{2-y}Se_2$ phase where these transformations are decoupled and are separated by about 50 K [22]. Indeed, persisting intensities of diffuse rods of the minor phase at 20 GPa, as compared to diminished intensities of the superstructure reflections of the main phase, support this possible scenario.

A question as to which phase could be the origin for the observed superconductivity in the SCII phases of related $A_xFe_{2-y}Se_2$ [18-20] should be addressed by taking into account experimental procedures during the corresponding $P$-dependent physical property measurements. As we have shown, compression at different temperatures results in different transformation kinetics and different phase compositions at analogues pressures. During the present synchrotron diffraction experiment pressure was scanned at selected, constant temperatures, and during the corresponding $P$-dependent physical properties measurements the pressure was increased at ambient temperature with following $T$-dependent scans [18-20]. Therefore, the measured superconducting signals at $P > 11.5$ GPa in $A_xFe_{2-y}Se_2$ originates from phases obtained at ambient temperature, possibly analogous to the HP $Cs_xFe_{2-y}Se_2$ phase synthesized by us under the same conditions - i.e. at room temperature.

**Conclusions**

We have synthesized a HP polymorph phase of $Cs_xFe_{2-y}Se_2$ for which the modification is characterized. Structurally the HP polymorph of $Cs_xFe_{2-y}Se_2$ is closely related to the LP minor superconducting phase of this system. Namely, their average structures correspond to the $I4/mmm$ $ThCr_2Si_2$-type arrangement without ordering within the deficient Fe-sublattice. In addition, the HP $Cs_xFe_{2-y}Se_2$ modification is less crystalline and nearly twice as soft as the corresponding main and minor parent phases. In addition, it is related to the $A_xFe_{2-y}Se_2$ phases with $T_c$ exceeding 40 K obtained with specific synthetic procedures, namely precise control of stoichiometry and annealing conditions, that do not feature iron vacancy ordering. The main difference between these materials are the differing synthetic routes: pressure- and temperature-mediated, respectively. The HP $Cs_xFe_{2-y}Se_2$ modification is formed at 11.8 GPa and, therefore, can be synthesized in quantities sufficient for physical property measurements using laboratory large volume presses. Once obtained, pure HP $Cs_xFe_{2-y}Se_2$ modification (and other $A_xFe_{2-y}Se_2$ phases in general) may exhibit improved superconducting properties at ambient pressure similar to the phases with superior superconducting properties synthesized by controlled cooling.